\documentclass[aps,prd,preprint,showpacs,amsmath,amssymb,floatfix]{revtex4}
\usepackage{epsfig}

\newcommand{\bq}{\begin{equation}}
\newcommand{\eq}{\end{equation}}
\newcommand{\bqa}{\begin{eqnarray}}
\newcommand{\eqa}{\end{eqnarray}}
\newcommand{\ben}{\begin{enumerate}}
\newcommand{\een}{\end{enumerate}}
\newcommand{\bc}{\begin{center}}
\newcommand{\ec}{\end{center}}
\newcommand{\bqb}{\begin{eqnarray*}}
\newcommand{\eqb}{\end{eqnarray*}}

\def\np#1#2#3{ Nucl. Phys. ${\bf{#1}}$:#2 (#3)}

\def\epj#1#2#3{ Eur. Phys. J. ${\bf{#1}}$:#2 (#3)}
\def\ijmp#1#2#3{ Int. J. Mod. Phys. ${\bf{#1}}$:#2 (#3)}

\global\nulldelimiterspace = 0pt

\begin{document}


\title{\vspace{0.1cm}
A peculiar property of SUSY 
amplitudes at high energy
}
\author{
M.~Beccaria$^{a,b}$,
F.M.~Renard$^c$
and C.~Verzegnassi$^{d,e}$ \\
\vspace{0.4cm}
}

\affiliation{\small
$^a$ $\mbox{Dipartimento di Fisica, Universit\`a del Salento, Italy}$ \\
\vspace{0.2cm}
$^b$ INFN, Sezione di Lecce, Italy\\
\vspace{0.2cm}
$^c$ Laboratoire de Physique Th\'{e}orique et Astroparticules,
Universit\'{e} Montpellier II, France\\
\vspace{0.2cm}
$^d$ $\mbox{Dipartimento di Fisica Teorica, Universit\`a di Trieste, Italy}$ \\
\vspace{0.2cm}
$^e$ INFN, Sezione di Trieste, Italy\\
}

\begin{abstract}
We observe that the electroweak one loop correction to the quark+gluon to quark+Higgs amplitude at high energy involves both single and quadratic logarithms of the energy 
in the SM case but only quadratic logarithms in the MSSM case. We explore the origin 
of this special SUSY cancellation, both in a diagrammatic way and through the 
splitting+Parameter Renormalization procedure. We show that it is not an accident but a remarkable and general SUSY property of the renormalized Higgs-fermion-fermion and Higgsino-sfermion-fermion vertices which directly reflects in such processes, for example in $bg\to tH^-$, $bg\to bH^0$, $bg\to bh^0$, $bg\to bA^0$, and through equivalence
in $bg\to tW^-_{long}$, $bg\to bZ_{long}$, as well as in $bg\to \tilde{t}\chi^-$, $bg\to\tilde{b}\chi^0$. This simplification of the high energy behaviour (which only 
leaves quadratic logarithms involving pure gauge couplings without any free parameter) allows to write simple relations among these various processes which could constitute genuine tests of the assumed SUSY model.\\
\end{abstract}

\pacs{12.15-y, 14.80.Ly}

\maketitle

\section{Introduction}
\label{sec:intro}

It is well-known that basic electroweak interactions reflect in a clear and simple way in the high energy logarithmic behaviour
of helicity amplitudes at one loop \cite{rsm1,rsm2,rmssm1,rmssm2}. To obtain this behaviour for a given process it is sufficient to use a table of coefficients corresponding to the splitting of the external particles and to the parameter renormalization (PR) corrections
to the coupling constants appearing in the Born terms~\cite{rmssm1,rmssm2} . These results have been checked by explicit one loop diagrammatic computations 
of $2\to 2$ processes in several cases \cite{proc1,proc2,proc3,proc4,proc5} and take the form:
\bqa 
F(\lambda_1, \lambda_2, \lambda_3, \lambda_4)
&\to& 
F^{Born}(\lambda_1, \lambda_2, \lambda_3, \lambda_4)
\left[1+{\alpha\over4\pi}\sum_{i=1,4} c(\lambda_i)\right] + 
\\
&& + \delta_{\rm PR} F^{Born}(\lambda_1, \lambda_2, \lambda_3, \lambda_4) ,
\nonumber
\eqa
where $\lambda_i$ are the particle helicities. The first correction $c(\lambda_i)$ is a contribution which
depends on the particular type of $i$-th particle. It is model dependent, but process independent. Its high energy 
structure is doubly logaritmic
\bqa
c(\lambda_i)=a(\lambda_i)\,L(s)+
b(\lambda_i)\,L^2(s)~~~~~~~~~~~~~~L(s)\equiv \log{s\over M^2},
\label{cab}
\eqa
where $\sqrt{s}$ is the c.m. energy and $M$ is a mass scale.
In the quadratic log part $M$ is either $M_W$ or $M_Z$ (see Appendix A). In the linear log part $M$ is an arbitrary reference mass and it is sometimes
convenient to take it as an average of the MSSM mass scales; a change of value just
corresponds to a modification of the constant term but does not affect the logarithmic
growth with the energy.

The second correction $\delta_{\rm PR}F^{Born}$ comes from the running of the tree level couplings and 
is a (single) logarithm of the energy related to the associated $\beta$ function. 

Along these analyses several differences betweeen the SM case and the MSSM case were already pointed out. They appear in the values of the splitting coefficients (both in gauge and in Yukawa terms) as well as in the PR coefficients (for instance the $\beta, \beta'$ electroweak RG functions). These differences are well identified in terms of the spectrum of SUSY particles which completes the SM 
spectrum.\\

These differences only concern the linear logarithmic parts (gauge and Yukawa), the quadratic logarithmic parts being fully controlled by the vector boson ($W^{\pm}$, $Z$, $\gamma$)
couplings which are identical in the SM and in the MSSM. In this linear 
logarithmic sector we have noticed a set of special features which had not been emphasized before and which could constitute a clear SUSY signature. This is the purpose of this short note.\\

We first noticed this specificity when studying the production of longitudinally polarized $W$ bosons in the 
process $bg\to tW^-$ \cite{bgtw}. The one loop correction to the Born amplitude for production of $W^-_{long}$ get linear and quadratic logarithms in the SM but the complete linear logarithms (of gauge and Yukawa origin) cancel when adding the SUSY contributions. We have checked this property both by explicit diagrammatic computation and through the splitting + PR method. 
In the longitudinal vector boson case a special SUSY property appears, leading, only in the MSSM, to the complete disappearence of the
linear logarithms in the full amplitude.

Notice that in the case of transverse boson production, things are quite different. For a general one loop process involving external $W_T$ bosons
we observe a cancellation between the $c(W_T)$ single logarithms and the $\beta$ function term associated to the gauge coupling.
In other words, 
\bqa
c^{splitt}(W^{\pm}_T)=a \,L(s)-~{1\over s^2_W}~\,L^2(s)
\label{csplw}
\eqa
but the PR of the gauge coupling $g$ gives
\bqa
{\delta g\over g}\equiv{\beta\over s^2_W}\,L(s)\equiv-a\,L(s).
\label{cprw}\eqa
The cancellation of single logs associated with $W_T$ legs just comes
from the fact that both terms ($W$ splitting and PR)
arise from the pure gauge coupling of the $W$ to fermions and sfermions.
This is peculiar to $W_T$ and leads to a cancellation which is valid both in SM and in the MSSM~\cite{rsm1,rsm2,rmssm1,rmssm2}.
This cancellation of the linear logarithmic occurs similarly for $W^0$ and $B$ vector bosons, and this has been checked by explicit diagrammatic computations in several cases
 \cite{proc3,proc4,proc5}.
Although interesting, this result is of little use.
Indeed, in a full process, like for instance $bg\to tW^-_{tr}$ one would need to add the single logarithms associated with the external $b$ and $t$ quark lines
and there would be no single logarithm cancellation in the total amplitude.

On the other hand, the full single logarithm cancellation that we discussed in the case of $bg\to tW_{long}$ works at the level of the full
MSSM one-loop amplitude and is therefore a SUSY related physical feature which, in principle, could be observable.

\medskip
In front of such a result we raise the following questions:
\begin{enumerate}
\item Is this cancellation an accident for this particular process
with its specific quantum numbers, or is it more general?

\item What are the basic SUSY properties leading to it? Is it one more smoothness aspect of SUSY?

\item Can this lead to specific SUSY tests?
\end{enumerate}

These are the 3 points that we successively develop in Sect.2,3,4 before concluding in Sect.5.

\section{Generality of this special cancellation}

To explore this point we first analyze in details the one loop contributions to the process $bg\to tG^-$ which is equivalent at high energy (at order $O(m^2/s)$) to  $bg\to tW^-_{long}$
but is much simpler to treat.\par
\begin{figure}[htb] 
\centering 
\epsfig{file=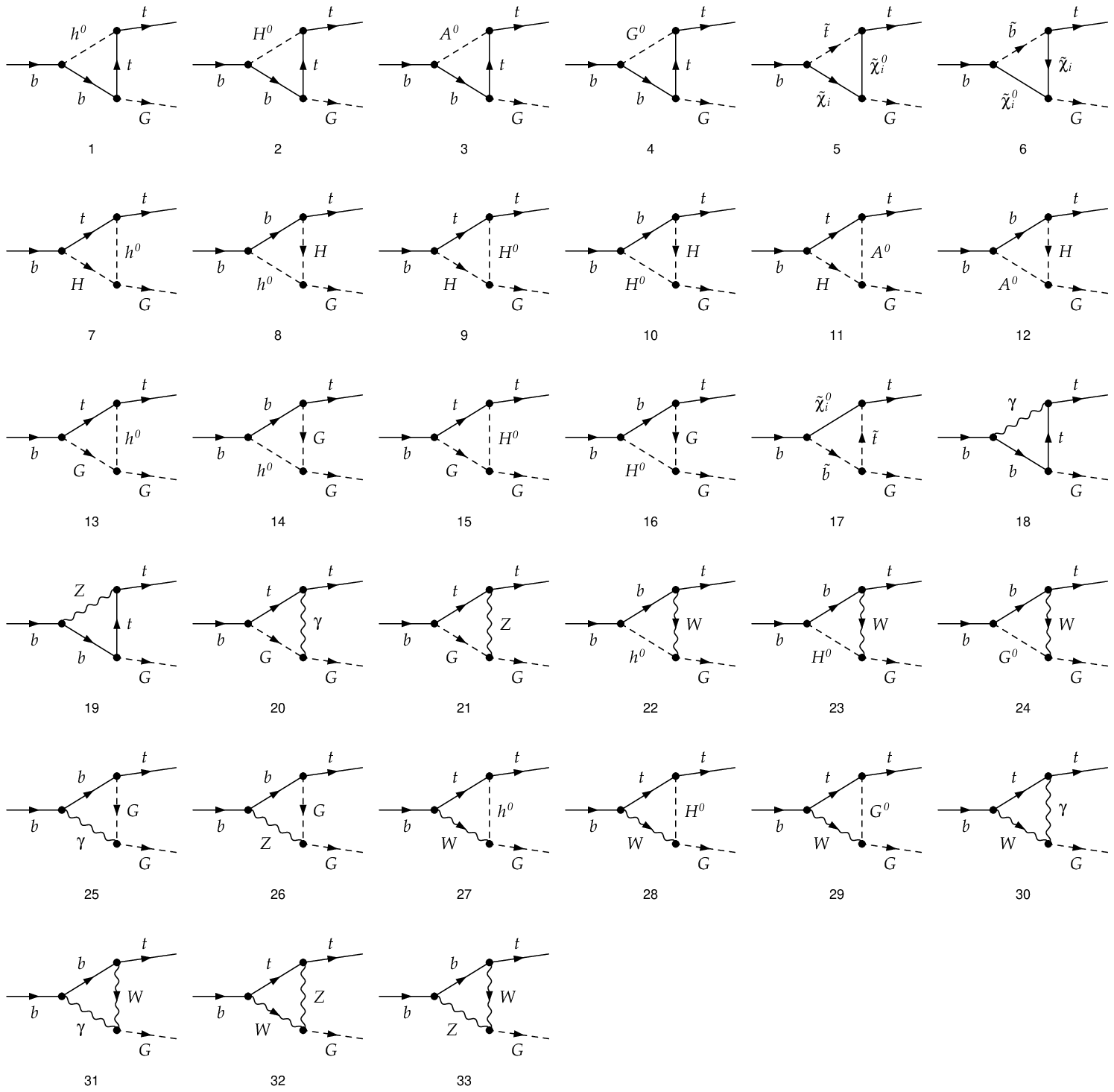, width=14cm, angle=0} 
\vspace{1.5cm} 
\caption{One loop triangle Feynman diagrams for the off-shell $b\to t\,G^-$.}
\label{fig:b2tG} 
\end{figure}

First, in a diagrammatic analysis we observe that a linear "gauge" logarithmic
contribution ${1+2c^2_W\over2s^2_Wc^2_W}\,L(s)$ arises from $(VSq)$, $(SVq)$,
$(qqVS)$ loops in the SM which then cancels with the additional $(\chi\chi\tilde{q})$ SUSY loop. 
The full list of triangle diagrams contributing the off-shell $b\to tG^-$ vertex is shown in Fig.~(\ref{fig:b2tG}).
In the Yukawa sector a linear logarithmic 
term $-~{m^2_a\over2s^2_WM^2_W}\,L(s)$ arises from the $(f'fH_{SM})$, $(f'fG^0)$ triangles in the SM, whereas in the the MSSM the THDM structure ($\Phi_1$ coupled to down quarks, $\Phi_2$ coupled to up quarks) leads to the separate cancellation of $(f'fH^0)+(f'fh^0)$
and of $(f'fG^0)+(f'fA^0)$ contributions.\par
We then look in details at the splitting +PR analysis of $bg\to tG^-$ for the $b_Lt_R$ amplitude (similar features appear for $b_Rt_L$).
The $b$ and $t$ splitting contributions are\\
in the SM ($L\equiv L(s)$)
\bqa
c(t_L)=c(b_L)={1+26c^2_W\over72s^2_Wc^2_W}[3\,L-L^2]
-{m^2_t+m^2_b\over8M^2_Ws^2_W}\,L
\label{ctlsm}\eqa
\bqa
c(t_R)={4\over18c^2_W}[3\,L-L^2]
-{m^2_t\over4M^2_Ws^2_W}\,L~~
c(b_R)={1\over18c^2_W}[3\,L-L^2]
-{m^2_b\over4M^2_Ws^2_W}\,L
\label{ctrsm}\eqa

and in the MSSM, with~~
$\tilde{m}^2_t=m^2_t(1+\cot^2\beta),~~
\tilde{m}^2_b=m^2_b(1+\tan^2\beta)
$~~
they are
\bqa
c(t_L)=c(b_L)={1+26c^2_W\over72s^2_Wc^2_W}[2\,L-L^2]
-2({\tilde{m}^2_t+\tilde{m}^2_b\over8M^2_Ws^2_W})\,L
\label{ctlmssm}\eqa
\bqa
c(t_R)={4\over18c^2_W}[2\,L-L^2]
-({\tilde{m}^2_t\over2M^2_Ws^2_W})\,L~~
c(b_R)={1\over18c^2_W}[2\,L-L^2]
-({\tilde{m}^2_b\over2M^2_Ws^2_W})\,L
\label{ctrmssm}\eqa

For $G^-$ splitting one has, in the SM

\bq
c(G^-)={1+2c^2_W\over8c^2_Ws^2_W}[4\,L-\,L^2]
-3{m^2_t+m^2_b\over4s^2_WM^2_W}\,L
\label{cgsm}\eq

and in the MSSM (with $G^-H^-$ mixing contribution)

\bq
c(G^-)={1+2c^2_W\over8c^2_Ws^2_W}[2\,L-\,L^2]
-3{\tilde{m}^2_t\over4s^2_WM^2_W}\,L
\label{cgmssm}\eq

The PR contribution arising from the $b_Lt_RG^-$ coupling is, in the SM

\bqa
&&({\sqrt{2}m_t\over v})_{SM}=
{gm_t\over\sqrt{2}M_W}
\to {\delta g\over g}+{\delta m_t\over m_t}
-{\delta M_W\over M_W}
\nonumber\\
&&\to \{-~{51+30c^2_W\over72s^2_Wc^2_W}+3{3m^2_t+m^2_b\over8s^2_WM^2_W}\}\,L
\label{cgprsm}\eqa
and in the MSSM
\bqa
&&({\sqrt{2}m_t\over v_2})_{MSSM}
\to {\delta g\over g}+{\delta m_t\over m_t}
-{\delta M_W\over M_W}-{\delta \sin\beta\over \sin\beta}
\nonumber\\
&&\to \{-~{26+28c^2_W\over36s^2_Wc^2_W}+{6\tilde{m}^2_t+\tilde{m}^2_b\over4s^2_WM^2_W}\}\,L
\label{cgprmssm}\eqa\\

\begin{figure}[htb] 
\centering 
\epsfig{file=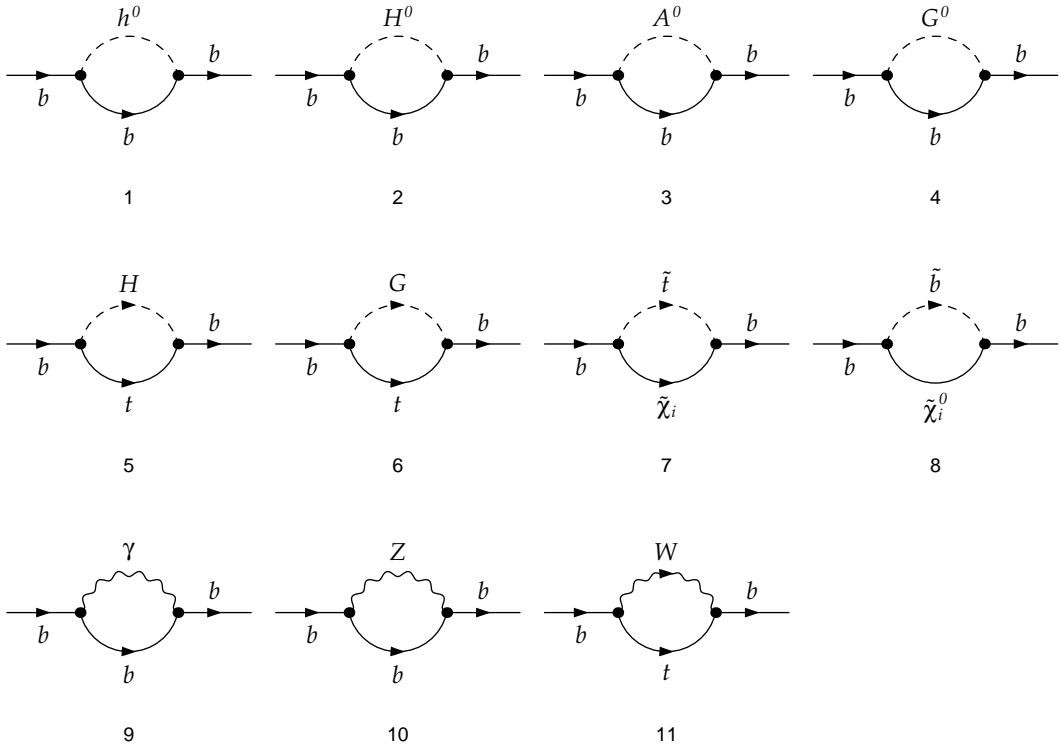, width=14cm, angle=0} 
\vspace{1.5cm} 
\caption{One loop self-energy of the $b$ quark.}
\label{fig:b2b} 
\end{figure}

Concerning the gauge part one observes that the $G^-$ splitting contribution cancels with the PR part $\delta g/g-\delta M_W/M_W$ separately in the SM and in the MSSM, whereas the $b,t$ splitting and the PR
part $\delta m_q/m_q$ combine to give ${1+2c^2_W\over2s^2_Wc^2_W}\,L$  in the SM but to cancel
in the MSSM due to the addition of the $(\chi\tilde{q})$ bubble contributions to the quark self-energy.
As an example, we show in Fig.~(\ref{fig:b2b}) the full list of diagrams contributing the $b$ quark self-energy.
For what concerns the Yukawa part, a similar property appears; $G^-$ splitting (including in MSSM the $G^-H^-$ mixing contribution)
cancels with $-\delta M_W/M_W$ (and $\delta \tan\beta/\tan\beta$ in the MSSM) whereas the $b,t$ splitting and the PR $\delta m_q/m_q$ combine to give   $-~{m^2_a\over2s^2_WM^2_W}\,L$ in the SM but also cancellation in the MSSM because of the THDM structure of the $(Hq)$ bubble contributions. The above SM residual term arises only from the scalar part $\Sigma_S$ of the quark self-energy and in MSSM it is cancelled by the additional SUSY contribution to $\Sigma_S$ in a way very similar to what happens in the diagrammatic analysis.\par
\begin{figure}[htb] 
\centering 
\epsfig{file=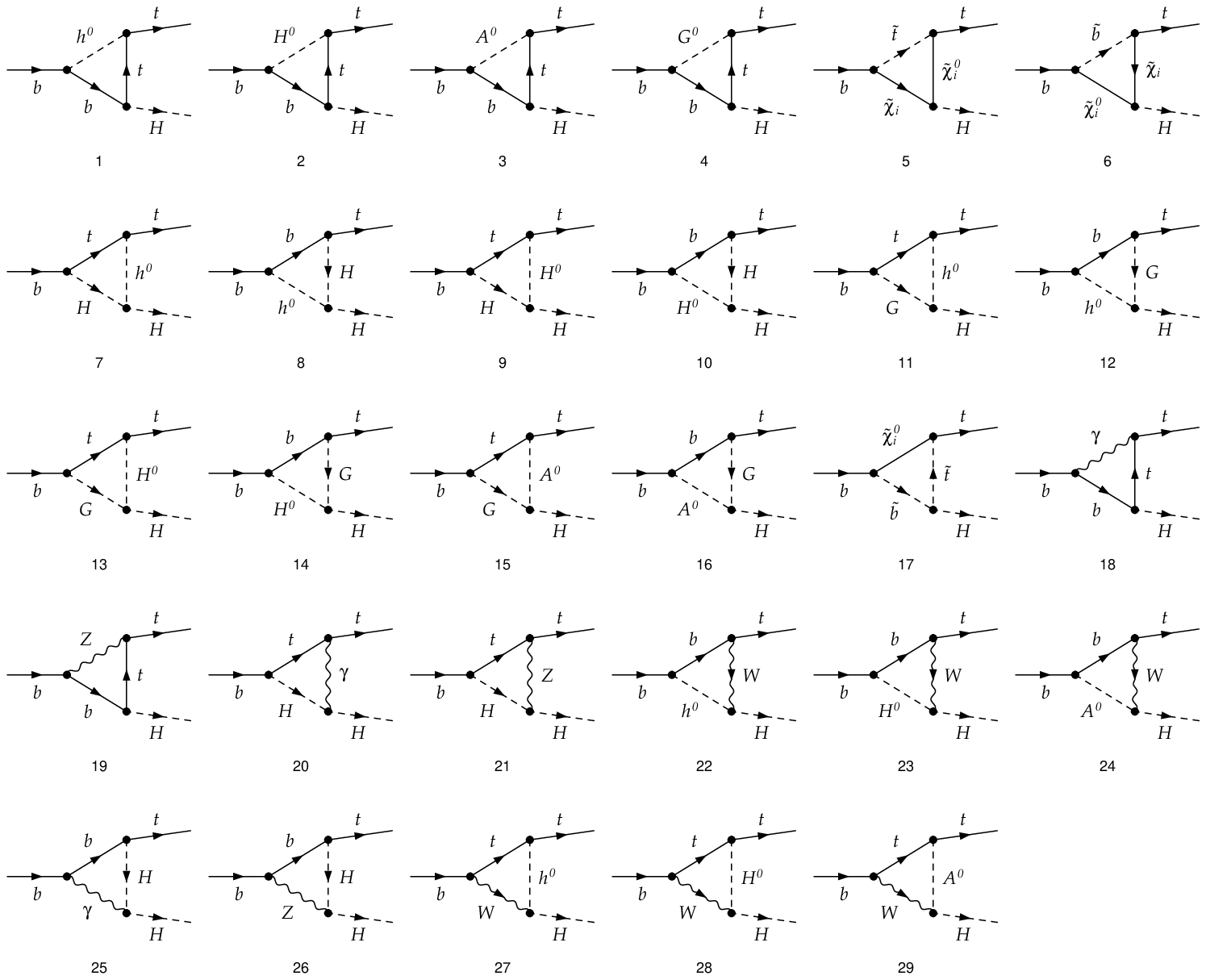, width=14cm, angle=0} 
\vspace{1.5cm} 
\caption{One loop triangle Feynman diagrams for the off-shell $b\to t\,H^-$.}
\label{fig:b2tH} 
\end{figure}

We then look at other processes. First we replace $W_{long}$ by $Z_{long}\simeq G^0$ and look at $bg\to bG^0$. We observe exactly the same properties. We then extend the same analysis to other types of Higgses.
In the SM case, $bg\to bH_{SM}$ gives the same resulting non zero residual terms as the pure SM $bg\to bG^0$.
In the MSSM $bg\to tH^-$, $bg\to bH^0$, $bg\to bh^0$, $bg\to bA^0$
behave similarly to $bg\to tG^-$ with the complete cancellation of the linear logarithms.
In each case the procedure of cancellation can be identified either  through the diagrams $(\chi\chi S)+(ffS)$ or in splitting+PR method through $(\chi\tilde{q})$ and $(Hq)$ contributions to the $\Sigma_S$ quark self-energy using the specific Higgs-quark-quark couplings and Higgs mixing.
In Fig.~(\ref{fig:b2tH}), we give the list of vertex diagrams for the case $b\to tH^-$.

\begin{figure}[htb] 
\centering 
\epsfig{file=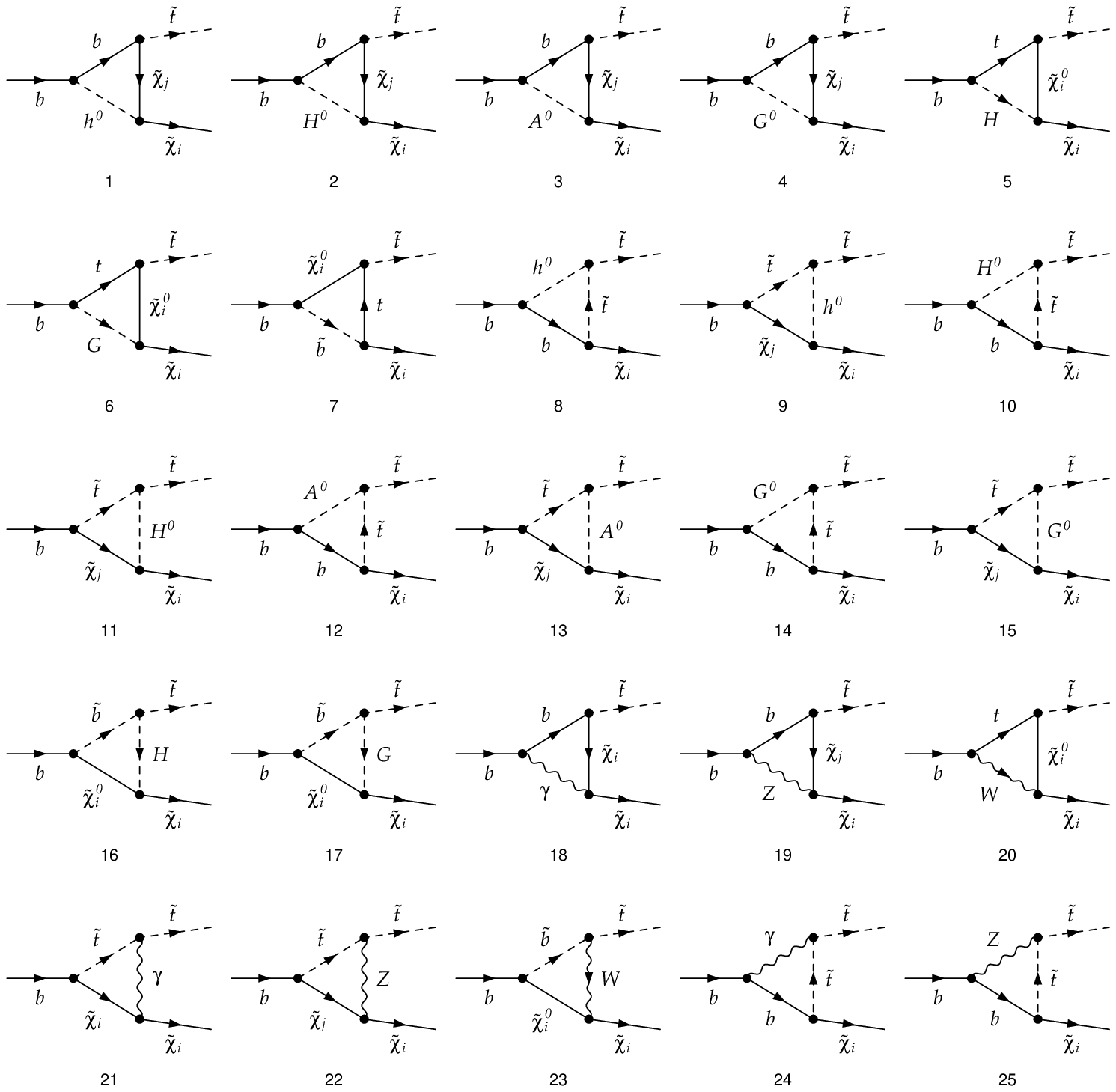, width=14cm, angle=0} 
\vspace{1.5cm} 
\caption{One loop triangle Feynman diagrams for the off-shell $b\to \widetilde{t}\,\chi^-$.}
\label{fig:b2stchi} 
\end{figure}

Driven by these SUSY considerations we made one more extension
by considering the production of SUSY partners, i.e. charginos and neutralinos replacing longitudinal gauge bosons or Higgs bosons, with illustration in 
the processes $bg\to \tilde{t}\chi^-$ (Fig.~(\ref{fig:b2stchi})) and $bg\to\tilde{b}\chi^0$. The diagrammatic and splittting+PR analyses of the processes show the same properties as for gauge and Higgs bosons, reflecting the supersymmetric invariance of this curiosity. 
The results are simple when separating the gaugino and the higgsino components. For the gaugino components i.e. factorizing out the Wino mixing element \cite{Rosiek} $Z^{\pm}_{1i}$ for charginos, or the Wino, Bino elements $Z^N_{2i}$, $Z^N_{1i}$ for neutralinos, one obtains the same pure quadratic logarithms coefficient for splitting+PR as for $W^{\pm},Z,\gamma$ and linear +quadratic logarithms coefficients for the associated quark and squark lines. For the Higgsino components,
i.e. factorizing the mixing element $Z^{\pm}_{21}$ for charginos or factorizing $Z^N_{3i,4i}$ for neutralinos, one observes the cancellation of the complete set of linear logarithms leaving an amplitude with pure quadratic logarithms. \\

\subsection{Connection with basic SUSY properties}

The above analyses have shown that this cancellation of the linear logarithms is not an accident but is a specific SUSY property of the $qqH$ and $\tilde{q}q\tilde{H}$ renormalized 
vertices which directly reflect in the $bg$ processes. For the gauge part the cancellation occurs due to the contribution of the spartners (gauginos, squarks) and for the 
Yukawa part due to the specific spectrum of the THDM. To relate this observed cancellation to some basic SUSY property, which is what one would naturally guess, is not simple. One plausible possibility would be to relate the cancellation of linear
logarithms to the non renormalization theorem of chiral vertices~\cite{nrth1,nrth2,nrth3}. However, this is not completely straightforward as is discussed 
for instance in~\cite{Kraus:2001kn}. Indeed, if perturbation theory is done in the Wess-Zumino gauge, then supersymmetry is non-linearly realized
and allows individual field renormalizations for all matter  fields ~\cite{Wess:1974jb}.
As a consequence chiral Green functions are
superficially convergent only up to  gauge-dependent field redefinitions. In the specific case of our calculation, 
one checks that the Yukawa contributions to the linear logarithm are exactly opposite to the UV divergence $\Delta$ ({\em i.e.} the combination $\Delta-L$ in the concerned diagrams).
These contributions are essentially {\em gaugeless} and the NR theorem applies in its simplest form. Thus the cancellation of the various 
$\Delta$ also leads to the cancellation of the overall coefficient multiplying $L$. This is not the case in SM, the $\Delta$ cancellation occuring only in the total sum (triangles and counter terms) 
with no special non-renormalization rule. For what concerns the gauge part the relation between cancellations and general SUSY properties is less obvious to us because of the above considerations and indeed the chiral vertex
is not convergent. Still, although we cannot honestly claim to have completely proved it, we shall regard the cancellation of linear logaritms in the complete amplitude as a gauge-invariant property, whose origin might be a consequence of the non renormalization theorem. We believe that a deeper investigation of this origin, certainly beyond the limits of this paper, would be motivated.

\section{Genuine tests of Supersymmetry}

This specific SUSY property of the $qqH$ and $\tilde{q}qH$ renormalized vertices could generate genuine tests of supersymmetry. 
In fact, a direct comparison of the energy dependence of the cross sections (logarithmic fits of the experimental results) for processes involving these vertices should confirm the absence of linear logarithmss.
With this purpose we list in Appendix A the explicit expressions of the one loop high energy amplitudes for a few typical cases. They may be used for comparison with experiments.\\

In addition one should note that these expressions contain only quadratic logarithms which involve no free parameter. All parameters are included in the Born terms. This allows to write simple relations among amplitudes and cross sections of several processes. They would constitute specific SUSY tests valid not only not only at the Born level, but also, at high energies, at 
one loop. In Appendix B we list them separately for
the charged sector ($bg\to tW^-$, $bg\to tH^-$, $bg\to \tilde{t}\chi^-_i$) and
for the neutral sector ($bg\to b\gamma$, $bg\to bZ$, $bg\to bH^{0}$, $bg\to bh^{0}$, $bg\to bA^{0}$, $bg\to\tilde{b}\chi^{0}_i$). These relations generalize the simpler ones written for the pure gauge/gaugino cases  
$ug\to dW$ and $ug\to\tilde{d}\chi^+$ in \cite{ugdw}.\\

Note that specific relations also appear among Higgs production processes.
The importance of the processes $bg\to b+$ Higgs has been for example emphasized in ref.\cite{camp}.
 As shown explicitely in Appendices A,B the amplitudes for Higgs production at Born level are related as
\bqa
&&{F^{H^-}_{-++}\over m_t\cot\beta}=
{F^{H^-}_{+--}\over m_b\tan\beta}=
-~{2\cos\beta\over m_b\cos\alpha}~F^{H^0}_{\pm\mp\mp}=
{2\cos\beta\over m_b\sin\alpha}~F^{h^0}_{\pm\mp\mp}\nonumber\\
&&
=
\mp{2i\over m_b\tan\beta} ~F^{A^0}_{\pm\mp\mp}
=\pm{2i\over m_b}~ F^{G^0}_{\pm\mp\mp}
\label{cfhm}\eqa

For the corresponding Born cross sections this would give:
\bqa
&&{1\over(m^2_t\cot^2\beta+m^2_b\tan^2\beta)}
~\sigma^{Born}(H^-)
={2\cos^2\beta\over m^2_b\cos^2\alpha}~\sigma^{Born}(H^0)
={2\cos^2\beta\over m^2_b\sin^2\alpha}~\sigma^{Born}(h^0)
\nonumber\\
&&
={2\over m^2_b\tan^2\beta}~\sigma^{Born}(A^0)
={2\over m^2_b}~\sigma^{Born}(G^0)
\label{csigborn}\eqa

But at one loop logarithmic level
$H^-$ production gets specific correction $C_{\mp\pm\pm}$ coefficients (see App.A),
different for $-++$ and $+--$ amplitudes, so that the first equality is violated.
However the 4 neutral productions get the same one loop high energy corrections
(see coefficients $N_{\mp\pm\pm}$ in App. A) and therefore the same leading high energy amplitudes so that the following
relations remain valid at this level:

\bqa
&&{\sigma(H^0)\over \cos^2\alpha}
={\sigma(h^0)\over \sin^2\alpha}
={\sigma(A^0)\over \sin^2\beta}
={\sigma(G^0)\over \cos^2\beta}
\label{ccsighn}\eqa
The relation with $\sigma(H^-)$ is more complicated due to the presence of the different coefficients $C_{\mp\pm\pm}$,
but it is well defined as these $N_{\mp\pm\pm}$ and $C_{\mp\pm\pm}$ coefficients only involve the parameters
$\alpha$ and $\beta$.\\
This whole set of relations among Higgs boson production cross sections could therefore provide the starting points of checks of the THDM structure. Of course, this proposal should take into account the specific experimental possibilities of LHC. In this respect, we feel that the following facts should be preliminary evidentiated:\\

1) The considered ratios of different Higgs rates have the remarkable property of
being independent of the involved parton distribution functions. The measurement of these rates
would represent a relatively clean test of the adopted supersymmetric
scheme.\\

2) The asymptotic expressions that we derived are expected to become valid
at large energies. This would be probably more realistic at a future high luminosity sLHC
collider~\cite{sLHC}. At LHC they might become relevant in the final
sector of the available cm energy, and be potentially visible in a
suitable final mass dependence of a  differential plot, rather than in the
total rate that would be affected by the lowest energy points.\\

3) Given the stressed relevance of the considered Higgs production
processes ratios, we feel that their complete one-loop electroweak (and
strong) calculation would be oportune. In this case, the clean  request on
the complete e.w. component of reproducing the simple logarithmic
expressions would provide a strong extra check of the validity of the
theoretical calculation. Also, it would allow to separate the low energy sector, theoretically more complicated
but provided by a certain numerical program, and join it with the predicted asymptotic expressions, in principle valid at the extreme machine energy sector.
This complete calculation is in fact already
being performed by our group\cite{bgth}.

\section{Conclusions and perspectives}

In conclusion we can say that
the peculiar feature that we discovered is one more example of the subtlety of SUSY
which adds to well-known and spectacular ones related to the non renormalization theorem. We have shown that it could lead to observable consequences. We have treated the simplest cases observable at LHC namely the various
$bg\to qH$ or $bg\to \tilde{q}\tilde{H}$ processes which directly reflect the
property of the $bqH$ or $b\tilde{q}\tilde{H}$ vertices. But other processes involving these vertices could be studied. Among them the simplest ones are for example $q\bar q\to VH$, $\tilde{V}\tilde{H}$, $VV$, $\chi\chi$. Experimental analyses of these processes at LHC or at a next proton-proton collider might constitute an alternative test of  some of the
assumed details of the involved Supersymmetric model.

\newpage

{\bf \large{Appendix A: High energy amplitudes at one loop}}\\

\underline{{\bf Process $bg\to tW^-_{tr}$.}}  Using the same notations as in ref.\cite{bgtw} the leading high energy helicity amplitudes are
(with the linear log terms [ln] denoted by $L$ only coming from $b$, $t$ lines):

\bqa
F^W_{-\pm-\pm}&\to&F^{Born}_{-\pm-\pm}[1+{\alpha\over4\pi}~C_{tr}]
\label{fwc}\eqa

\bqa
F^{Born}_{-+-+}\to~{eg_s\over\sqrt{2}s_W}({\lambda^l\over2})
2\cos{\theta\over2}~~~~~~
F^{Born}_{----}\to~{eg_s\over\sqrt{2}s_W}({\lambda^l\over2})
({2\over\cos{\theta\over2}})
\label{fwborn}\eqa

\bqa
C_{tr}&=&{1+26c^2_W\over18s^2_Wc^2_W}
~L-[{m^2_t\over2s^2_WM^2_W}(1+\cot^2\beta)+
{m^2_b\over2s^2_WM^2_W}(1+\tan^2\beta)]L\nonumber\\
&&
-\{
{1\over2s^2_W}[ln^2{-u\over m^2_W}+ln^2{-u\over m^2_Z}]
+{1-10c^2_W\over36s^2_Wc^2_W}~ln^2{-t\over m^2_Z}~\}
\label{cwtr}\eqa \\

{\underline{\bf Process $bg\to tW^-_{long}$.}}   The leading amplitudes are (with no [ln] at all).

\bqa
F^W_{\mp,\pm,\pm,0}=F^{Born}_{\mp,\pm,\pm,0}[1+{\alpha\over4\pi}C_{\mp,\pm,\pm}]
\label{fgc}\eqa

with
\bqa
F^{Born}_{-,+,+,0}=eg_s({\lambda^l\over2})
{m_t\over s_WM_W}
\cos{\theta\over2}({1-\cos\theta\over1+\cos\theta})
~~F^{Born}_{+,-,-,0}=eg_s({\lambda^l\over2})
{m_b\over s_WM_W}
\cos{\theta\over2}({1-\cos\theta\over1+\cos\theta})
\label{fgborn}\eqa

\bqa
&&C_{-,+,+}=
-{1\over3c^2_W}\log^2\frac{s}{m^2_Z}
-{1\over9c^2_W}\log^2\frac{-t}{m^2_W}
+{1-4c^2_W\over12s^2_W c^2_W}\log^2\frac{-u}{M_Z^2}
-~{1\over2s^2_W}\log^2\frac{-u}{M_W^2}~~
\label{cgmpp}\eqa 
\bqa
&&C_{+,-,-}=
-{1+2c^2_W\over12s^2_Wc^2_W}\log^2\frac{s}{m^2_Z}
-{1\over2s^2_W}\log^2\frac{s}{m^2_W}
+{1\over18c^2_W}\log^2\frac{-t}{M_W^2}
-~{1\over6s^2_W}\log^2\frac{-u}{M_W^2}~~
\label{cgpmm}\eqa\\

\underline{{\bf Process $bg\to tH^-$.}}  The amplitudes are expressed in terms of the same $C_{\mp,\pm,\pm}$ coefficients as in the previous $W_{long}$ case
\bqa
F^{H^-}_{\mp,\pm,\pm}&=&F^{Born}_{\mp,\pm,\pm}\{1+[{\alpha\over4\pi}]C_{\mp,\pm,\pm}\}
\label{fhmc}\eqa
\bqa
F^{Born}_{-,+,+}=-eg_s({\lambda^l\over2})
{m_t\over s_WM_W}
\cos{\theta\over2}({1-\cos\theta\over1+\cos\theta})\cot\beta
\label{fhmbornmpp}\eqa
\bqa
F^{Born}_{+,-,-}=eg_s({\lambda^l\over2})
{m_b\over s_WM_W}
\cos{\theta\over2}({1-\cos\theta\over1+\cos\theta})\tan\beta
\label{fhmbornpmm}\eqa

\underline{{\bf Process $bg\to \tilde{t}_a\chi^-_i$.}} 
It is convenient to consider separately the gaugino and the higgsino components,
using also the $\tilde{t}_L$ and $\tilde{t}_R$ decomposition.
The gaugino component has only $[ln]$ from $b$, $\tilde{t}$ lines
\bqa
F^{\chi}_{-++}(\tilde{t}_L)=F^{Born}_{-++}(\tilde{t}_L)\{1+{\alpha\over4\pi}C_{tr}\}
\label{fchimc}\eqa

with the same $C_{tr}$ as in the $W_{tr}$ case and 
\bqa
&&F^{Born}_{-++}(\tilde{t}_L)=-g_s({\lambda^l\over2})\sqrt{2}
A^L_i(\tilde{t}_L)\sin{\theta\over2}
~~~~~~~A^L_i(\tilde{t}_L)=-~{e\over s_W}Z^+_{1i}
\label{fchimgborn}\eqa

As in the $W_{long}$ case  the higgsino components have no $[ln]$ at all :
\bqa
F^{\chi}_{-++}(\tilde{t}_R)=F^{Born}_{-++}(\tilde{t}_R)[1+{\alpha\over4\pi}C_{-++}]~~~~~
F^{\chi}_{+--}(\tilde{t}_L)=F^{Born}_{+--}(\tilde{t}_L)[1+{\alpha\over4\pi}C_{+--}]
\label{fchimhc}\eqa

\bqa
F^{Born}_{-++}(\tilde{t}_R)=-~{\alpha\over4\pi}g_s({\lambda^l\over2})\sqrt{2}
A^L_i(\tilde{t}_R)\sin{\theta\over2}
~~~~F^{Born}_{+--}(\tilde{t}_L)=g_s({\lambda^l\over2})\sqrt{2}
A^R_i(\tilde{t}_L)\sin{\theta\over2}
\label{fchimhborn}\eqa 

\bq
A^L_i(\tilde{t}_R)={e m_t\over \sqrt{2}M_Ws_Wsin\beta}Z^+_{2i}  ~~~~~~~~A^R_i(\tilde{t}_L)={e m_b\over \sqrt{2}M_Ws_Wcos\beta}Z^{-*}_{2i}
\label{ai}\eq

\newpage

\underline{{\bf Processes $bg\to b\gamma$
and  $bg\to bZ_{tr}$}}. 
For future comparisons we separate the $B^0$ and $W^0$
components of $\gamma, Z$

\bq
F^{\gamma}_{\mp,\mu,\mp,\mu}=c_WF^{B0}_{\mp,\mu,\mp,\mu}+s_WF^{W0}_{\mp,\mu,\mp,\mu}
~~~F^{Z}_{\mp,\mu,\mp,\mu}=c_WF^{W^0}_{\mp,\mu,\mp,\mu}-s_WF^{B0}_{\mp,\mu,\mp,\mu}
\label{fbw}\eq

with 

\bq
F^{B^0~Born}_{-+-+}=~{eg_s\over6c_W}({\lambda^l\over2})(2\cos{\theta\over2})
~~~~~
F^{B^0~Born}_{----}=~{eg_s\over6c_W}({\lambda^l\over2})({2\over\cos{\theta\over2}})
\label{fbwborn}\eq

\bq
F^{B^0~Born}_{++++}=-~{eg_s\over3c_W}({\lambda^l\over2})(2\cos{\theta\over2})
~~~~~
F^{B^0~Born}_{+-+-}=-~{eg_s\over3c_W}({\lambda^l\over2})({2\over\cos{\theta\over2}})
\label{fbzborn}\eq

\bq
F^{W^0~Born}_{-+-+}=~{eg_s\over2s_W}({\lambda^l\over2})(2\cos{\theta\over2})
~~~~~
F^{W^0~Born}_{----}=~{eg_s\over2s_W}({\lambda^l\over2})({2\over\cos{\theta\over2}})
\label{fwzborn}\eq

and

\bq
F^{B^0,W^0}_{-,\mu,-,\mu}=F^{B^0,W^0~Born}_{-,\mu,-,\mu}[1+{\alpha\over4\pi}C^{B^0,W^0,Born}_{tr,-}]
\label{fbwzcm}\eq

\bq
F^{B^0}_{+,\mu,+,\mu}=F^{B^0~Born}_{+,\mu,+,\mu}[1+{\alpha\over4\pi}C^{B^0,Born}_{tr,+}]
\label{fbwzcp}\eq

\bq
C^{B^0,Born}_{tr,-}={1+26c^2_W\over18s^2_Wc^2_W}L
-{\tilde{m}^2_t+\tilde{m}^2_b\over2s^2_WM^2_W}L
-[{1+8c^2_W\over36s^2_Wc^2_W}]ln^2t_Z
-[{1\over s^2_W}]ln^2t_W
\label{cbzbornm}\eq

\bq
C^{B^0,Born}_{tr,+}={1\over9c^2_W}[2L-ln^2t_Z]
-{\tilde{m}^2_b\over s^2_WM^2_W}[ln]
\label{cbzbornp}\eq

\bqa
&&C^{W^0,Born}_{tr,-}={1+26c^2_W\over18s^2_Wc^2_W}L
-{\tilde{m}^2_t+\tilde{m}^2_b\over2s^2_WM^2_W}L
-{1+8c^2_W\over36s^2_Wc^2_W}ln^2t_Z
+~{3-4s^2_W\over2s^2_W(3-2s^2_W)}ln^2t_W\nonumber\\
&&
-{3c^2_W\over2s^2_W(3-2s^2_W)}[ln^2s_Z+ln^2s_W
+ln^2u_Z+ln^2u_W]
\label{cwzborn}\eqa

and one sees that $[ln]$ only arise from $b$ lines.\\

\underline{{\bf Process   $bg\to bZ_{long}$}}. The leading amplitudes involves also only $ln^2$ terms:
\bqa
F_{-++0}=
F^{Born}_{-++0}
\{1+{\alpha\over4\pi}N_{-++}\}
\label{fzlmpp}\eqa
\bqa
F_{+--0}&=&
F^{Born}_{+--0}
\{1+{\alpha\over4\pi}N_{+--}\}
\label{fzlbornp}\eqa
\bq
F^{Born}_{-++0}\to -F^{Born}_{+--0}\to
-~{eg_s\over2s_Wc_W}({\lambda^l\over2})
{\sqrt{2}m_b\over M_Z}
\cos{\theta\over2}({1-\cos\theta\over1+\cos\theta})
\label{fzlbornm}\eq
\bqa
N_{-++}&=&
-~{1\over6c^2_W}ln^2s_Z+~{1\over18c^2_W}ln^2t_Z
-~{1+2c^2_W\over12c^2_Ws^2_W}ln^2u_Z
-~{1\over2s^2_W}ln^2u_W
\label{cnmpp}\eqa
\bqa
N_{+--}&=&
-~{1+2c^2_W\over12c^2_Ws^2_W}ln^2s_Z
-~{1\over2s^2_W}ln^2s_W
+~{1\over18c^2_W}ln^2t_Z
-~{1\over6s^2_W}ln^2u_Z
\label{cnpmm}\eqa

\underline{{\bf Processes $bg\to bH^0,bh^0,bA^0,bG^0$}}. The amplitudes are given in terms of the same $N_{\mp\pm\pm}$ as in the above $Z_{long}$ case (with no $[ln]$ at all)
\bqa
F^H_{\mp,\pm,\pm}&=&F^{Born}_{\mp,\pm,\pm}
[1+{\alpha\over4\pi}N_{\mp\pm\pm}]
\label{fhnn}\eqa
\bqa
F^{Born}_{\mp,\pm,\pm}\to
-~\sqrt{2}c^{L,R}g_s({\lambda^l\over2})
\cos{\theta\over2}
({1-\cos\theta\over1+\cos\theta})
\label{fhnborn}\eqa 
\bqa
c^L_{H^0b}=c^R_{H^0b}=
-~({em_b\over2s_WM_W}){cos\alpha\over\cos\beta}
~~~~~~~
c^L_{h^0b}=c^R_{h^0b}
=({em_b\over2s_WM_W}){sin\alpha\over\cos\beta}
\label{chnh}\eqa
\bqa
c^L_{A^0b}=c^{R*}_{A^0b}=(-i)({em_b\over2s_WM_W})tan\beta~~~~~~
c^L_{G^0b}=c^{R*}_{G^0b}=~(i)({em_b\over2s_WM_W})
\label{chna}\eqa
One can check the equivalence of $G^0$ with $Z_{long}$.\\

\newpage

\underline{{\bf Process $bg\to \tilde{b}\chi^0_i$}}.
Separating the gaugino (Wino, Bino) and the higgsino components and using the $\tilde{b}_L$, $\tilde{b}_R$ decomposition we obtain:\\

\underline{for the
gaugino parts} (with the same coefficients as in $\gamma,Z$ and only $[ln]$ from $b, \tilde{b}$ lines)

\bqa
F_{+--}(\tilde{b}_{R})\to F^{Born~Bino}_{+--}(\tilde{b}_{R})
\{1+ 
{\alpha\over4\pi}C^{B0}_{tr,+}\}
\label{fchizp}\eqa

\bqa
F_{-++}(\tilde{b}_{L})\to 
F^{Born~Bino}_{-++}(\tilde{b}_{L})
\{1+ 
{\alpha\over4\pi}C^{B0}_{tr,-}\}
+F^{Born~Wino}_{-++}(\tilde{b}_{L})
\{1+ 
{\alpha\over4\pi}C^{W0}_{tr,-}\}
\label{fchizm}\eqa

\bqa
F^{Born~Bino}_{-++}(\tilde{b}_{L}) 
={eg_sZ^N_{1i}\over3c_W}({\lambda^l\over2})\sin{\theta\over2}
~~~~~
F^{Born~Wino}_{-++}(\tilde{b}_{L}) 
=-~{eg_sZ^N_{2i}\over s_W}({\lambda^l\over2})\sin{\theta\over2}
\label{fchizborn}\eqa

\underline{For the higgsino parts} (with  the same coefficients as in the $Z_{long}$ case and no $[ln]$ at all) 
\bqa
&&F_{+--}(\tilde{b}_{L})\to F^{Born~Higg}_{+--}(\tilde{b}_{L})
\{1+ 
{\alpha\over4\pi}N_{+--}\}\nonumber\\
&&
F_{-++}(\tilde{b}_{R})\to 
F^{Born~Higg}_{-++}(\tilde{b}_{R})
\{1+ 
{\alpha\over4\pi}N_{-++}\}
\label{fchihig}\eqa

\bqa
F^{Born~Higg}_{+--}(\tilde{b}_{L})=F^{Born~Higg~*}_{-++}(\tilde{b}_{R}) 
=-~{eg_sm_b\over M_Ws_W\cos\beta}Z^{N*}_{3i}({\lambda^l\over2})\sin{\theta\over2}
\label{fchihigborn}\eqa

\newpage

{\bf \large{Appendix B: SUSY relations}}\\

\underline{In the charged sector} $W^{\pm},H^{\pm},\chi^{\pm}$, looking at the expressions of Appendix A for the high energy amplitudes and using in particular the fact that the one loop corrections for longitudinal gauge bosons and Higgses involve only squared logs without any free parameter, one obtains the relations
\bqa
\cot{\theta\over2}F^{\chi}_{-++}(\tilde{t}_L)/Z^+_{1i}=
-F^W_{-+-+}=-\cos^2{\theta\over2}~F^W_{----}
\label{relfw}\eqa
\bqa
F^{\chi}_{-++}(\tilde{t}_R)/Z^+_{2i}=-\cot{\theta\over2}~F^W_{-++0}
/\sin\beta=\cot{\theta\over2}~F^{H^-}_{-++}/\cos\beta
\label{relfhm}\eqa
\bqa
F^{\chi}_{+--}(\tilde{t}_L)/Z^{-*}_{2i}=\cot{\theta\over2}~F^W_{+--0}
/\cos\beta=-\cot{\theta\over2}~F^{H^-}_{+--}/\sin\beta
\label{relfmhp}\eqa

For polarized cross sections one gets
\bqa
&&\sum_i{d\sigma(bg \to\chi^{-}_i+\tilde{t_{L}})_{-++}
\over dcos\theta}
=({ut\over u^2+s^2})
{d\sigma(bg \to t+W^{-}_T )\over dcos\theta}
\label{sigwt}\eqa
\bqa
\sum_i{d\sigma(bg 
\to\chi^{-}_i+\tilde{t_{R}})_{-++}\over dcos\theta}
=({u\over t})
[{d\sigma(bg \to t+W^{-}_{long} )_{-++0}\over dcos\theta}
+{d\sigma(bg \to t+H^{-} )_{-++}\over dcos\theta}]
\label{sigwl}\eqa
\bqa
\sum_i{d\sigma(bg 
\to\chi^{-}_i+\tilde{t_{L}})_{+--}\over dcos\theta}
=({u\over t})
[{d\sigma(bg \to t+W^{-}_{long} )_{+--0}\over dcos\theta}
+{d\sigma(bg \to t+H^{-} )_{+--}\over dcos\theta}]
\label{sigwtl}\eqa

and globally:

\bqa
&&\sum_i[{d\sigma(bg \to\chi^{-}_i+\tilde{t_{L}})\over dcos\theta}
+{d\sigma(bg \to\chi^{-}_i+\tilde{t_{R}})\over dcos\theta}]
=({ut\over u^2+s^2})
~{d\sigma(bg \to t+W^{-}_T )\over dcos\theta}\nonumber\\
&&
+({u\over t})
[{d\sigma(bg \to t+H^- )\over dcos\theta}
+{d\sigma(bg \to t+W^-_{long})\over dcos\theta}]
\label{sighw}\eqa

\newpage

\underline{In the neutral sector $\gamma, Z,H^0, h^0, A^0, \chi^0$ }

The relations among gaugino amplitudes are 
\bqa
F^{Wino}_{-++}(\tilde{b}_{L})/Z^N_{2i}=
\tan{\theta\over2}~F^W_{-+-+}
=\tan{\theta\over2}\cos^2{\theta\over2}~F^W_{----}
\label{relfwino}\eqa
\bqa
F^{Bino}_{-++}(\tilde{b}_{L})/Z^N_{1i}=
\tan{\theta\over2}~F^B_{-+-+}
=\tan{\theta\over2}\cos^2{\theta\over2}~F^B_{----}
\label{relfbinom}\eqa
\bqa
F^{Bino}_{+--}(\tilde{b}_{R})/Z^{N*}_{1i}=
\tan{\theta\over2}~F^B_{+-+-}
=\tan{\theta\over2}\cos^2{\theta\over2}~F^B_{++++}
\label{relfbinop}\eqa
\bqa
F^{Wino}_{+--}(\tilde{b}_{R})=F^W_{+,\mu,+,\mu}=0
\label{relwinop}\eqa

and among Higgsino  amplitudes
\bqa
F^{\chi}_{+--}(\tilde{b}_{L})=-{A^{0R}_i(\tilde{b}_{L})\over c^R_H}
\cot{\theta\over2}~F^H_{+--}
\label{relfchihigp}\eqa
\bqa
F^{\chi}_{-++}(\tilde{b}_{R})={A^{0L}_i(\tilde{b}_{R})\over c^L_H}
\cot{\theta\over2}~F^H_{-++}
\label{relfchihigm}\eqa
\noindent
They are valid for any $bH$ final state using the appropriate $H$
coupling $c^{L,R}$ given in Appendix A and
\bqa
A^{0L}_i(\tilde{b}_{R})=-~{em_b\over \sqrt{2}s_WM_W\cos\beta}
Z^N_{3i}
~~~~~
A^{0R}_i(\tilde{b}_{L})=-~{em_b\over \sqrt{2}s_WM_W\cos\beta}
Z^{N*}_{3i}
\label{a0i}\eqa

First notice the relations among the Higgs production
amplitudes \underline{at Born level}
\bqa
&&{F^{H^-}_{-++}\over m_t\cot\beta}=
{F^{H^-}_{+--}\over m_b\tan\beta}=
-~{2\cos\beta\over m_b\cos\alpha}~F^{H^0}_{\pm\mp\mp}=
{2\cos\beta\over m_b\sin\alpha}~F^{h^0}_{\pm\mp\mp}\nonumber\\
&&=
\mp{2i\over m_b\tan\beta} ~F^{A^0}_{\pm\mp\mp}
=\pm{2i\over m_b}~ F^{G^0}_{\pm\mp\mp}
\label{relfhigg}\eqa

For the Born cross section this would give:
\bqa
&&{1\over(m^2_t\cot^2\beta+m^2_b\tan^2\beta)}
~\sigma^{Born}(H^-)
={2\cos^2\beta\over m^2_b\cos^2\alpha}~\sigma^{Born}(H^0)
={2\cos^2\beta\over m^2_b\sin^2\alpha}~\sigma^{Born}(h^0)
\nonumber\\
&&
={2\over m^2_b\tan^2\beta}~\sigma^{Born}(A^0)
={2\over m^2_b}~\sigma^{Born}(G^0)
\label{relfhiggborn}\eqa

But at one loop logarithmic level
$H^-$ production gets specific correction coefficients,
different for $-++$ and $+--$ amplitudes, such that the first equality is violated.
However the 4 neutral productions get the same one loop high energy corrections
(see above) and the same leading amplitudes such that the following
relations remain valid at this level:

\bqa
&&{\sigma(H^0)\over \cos^2\alpha}
={\sigma(h^0)\over \sin^2\alpha}
={\sigma(A^0)\over \sin^2\beta}
={\sigma(G^0)\over \cos^2\beta}
\label{relsign}\eqa

Secondly we can relate the neutralino production cross sections to those of $\gamma, Z$ and Higgs production.
Eq() gives directly these relations for the pure Bino, Wino and Higgsino cases. The cross section $\sigma_i$ for physical
neutralino ($i=1,4$) production are then given by

\bqa
\sigma_i=\sigma(Bino)|Z^N_{1i}|^2+\sigma(Wino)|Z^N_{2i}|^2+\sigma(Higgsino)|Z^N_{3i}|^2
\label{sigibwhigg}\eqa

We refrain to write the obvious but lengthy expressions of the  Bino, Wino and Higgsino cross sections in terms of these physical cross sections by solving the above equation. Note nevertheless that there is no $Z^N_{4i}$ contribution in the processes $bg \to \tilde{b}\chi^0$ because in the THDM structure this 4th component only appear in the top quark sector. This implies one constraint among the set of physical cross sections. We just write the global relation obtained by using the orthogonality $\Sigma_i|Z^N_{ji}|^2=1$

\bqa
&&\sum_i[{d\sigma(bg \to\chi^{0}_i+\tilde{b_{L}})\over dcos\theta}
+{d\sigma(bg \to\chi^{0}_i+\tilde{b_{R}})\over dcos\theta}]
=({ut\over u^2+s^2})~
[{d\sigma(bg \to b\gamma)\over dcos\theta}+
~{d\sigma(bg \to b+Z_T )\over dcos\theta}]\nonumber\\
&&
+({u\over t})
[{d\sigma(bg \to b+H^0 )\over dcos\theta}+
{d\sigma(bg \to b+h^0 )\over dcos\theta}
+
{d\sigma(bg \to b+A^0 )\over dcos\theta}
+{d\sigma(bg \to t+Z_{long})\over dcos\theta}]
\label{relsigchiz}\eqa

with 
\bqa
&&~~~~{d\sigma(bg \to b+H^0 )\over dcos\theta}+
{d\sigma(bg \to b+h^0 )\over dcos\theta}
=
{d\sigma(bg \to b+A^0 )\over dcos\theta}
+{d\sigma(bg \to t+Z_{long})\over dcos\theta}~~~~~~
\label{relhhag}~~~~~~\eqa

\newpage

\newpage

\newpage

\newpage

\begin{thebibliography}{99}

\bibitem{rmssm1}  M. Beccaria,
F.M. Renard and C. Verzegnassi, hep-ph/0203254;
"Logarithmic Fingerprints of Virtual Supersymmetry"
Linear Collider note LC-TH-2002-005,  GDR Supersymmetrie
note GDR-S-081.

\bibitem{rmssm2}  
M. Beccaria, M. Melles, F. M. Renard,
S. Trimarchi, C. Verzegnassi, \ijmp{A18}{5069}{2003};
hep-ph/0304110.

\bibitem{rsm1}  A. Denner and S. Pozzorini,
\epj{C18}{461}{2001}.

\bibitem{rsm2}
A. Denner, B. Jantzen and S. Pozzorini,
\np{B761}{1}{2007},   hep-ph/0608326.


\bibitem{proc1} M. Beccaria, H. Eberl, F.M. Renard and C. Verzegnassi,
hep-ph/0406253, Phys. Rev. D 70, 071301(R) (2004).

\bibitem{proc2} 
M. Beccaria, F.M. Renard and C. Verzegnassi,
hep-ph/0410089; Phys.Rev.D71,033005,2005.

\bibitem{proc3} 
M.~Beccaria,  G.~Macorini, L.~Panizzi, F.M.~Renard, C.~Verzegnassi,
hep-ph/0710.5357.

\bibitem{proc4} G.J. Gounaris, J. Layssac, F.M. Renard,
Phys.Rev.D67(2003)013012.

\bibitem{proc5} M. Beccaria, F.M. Renard, C. Verzegnassi,
hep-ph/0304175; Nucl.Phys. B663(2003)394. 

\bibitem{bgtw} 
M. Beccaria, G. Macorini, F.M. Renard, C. Verzegnassi,
hep-ph/0601175; Phys.Rev.D 73:093001(2006)


\bibitem{Rosiek} J. Rosiek, Phys.Rev.D41,3464(1990).

\bibitem{nrth1} 
 K.~Fujikawa and W.~Lang,
 Nucl.\ Phys.\  B {\bf 88}, 61 (1975).

\bibitem{nrth2} 
 M.~T.~Grisaru, W.~Siegel and M.~Rocek,
 Nucl.\ Phys.\  B {\bf 159}, 429 (1979).

\bibitem{nrth3} 
 M.~T.~Grisaru and W.~Siegel,
 Nucl.\ Phys.\  B {\bf 201}, 292 (1982)
 [Erratum-ibid.\  B {\bf 206}, 496 (1982)].

\bibitem{Kraus:2001kn}
 E.~Kraus and D.~Stockinger,
 Nucl.\ Phys.\  B {\bf 626}, 73 (2002)
 [arXiv:hep-th/0105028].


\bibitem{Wess:1974jb}
 J.~Wess and B.~Zumino,
 Nucl.\ Phys.\  B {\bf 78}, 1 (1974).

\bibitem{ugdw} G.J. Gounaris, J. Layssac, F.M. Renard,
Phys.Rev.D77:093007,2008.

\bibitem{camp} J. Campbell, R.K. Ellis, F. Maltoni, S. Willenbrock, Phys.Rev.D67:095002,2003, [arXiv:hep-ph/0204093].


\bibitem{bgth} M.~Beccaria,  G.~Macorini, L.~Panizzi, F.M.~Renard, C.~Verzegnassi, in progress.

\bibitem{sLHC} {\tt http://atlas.web.cern.ch/Atlas/GROUPS/UPGRADES/}


\end{thebibliography}
\end{document}